\documentclass[%
 reprint,
 amsmath,amssymb,
 aps,
]{revtex4-1}
\usepackage[english]{babel}
\usepackage[dvips]{graphicx}
\usepackage{dcolumn}
\usepackage{bm}

\begin{document}

\preprint{APS/123-QED}

\title{Monte Carlo Simulation of Melting and Lattice Relaxation of the (111) Surface of Silver }
\author{Virgile Bocchetti}
 \email{virgile.bocchetti@u-cergy.fr}

\affiliation{Laboratoire de Physique Th\'eorique et Mod\'elisation, Universit\'e de Cergy-Pontoise, CNRS, UMR 8089\\
2, Avenue Adolphe Chauvin, 95302 Cergy-Pontoise Cedex,  France}

\author{H. T. Diep}%
 \email{diep@u-cergy.fr, corresponding author}
\affiliation{Laboratoire de Physique Th\'eorique et Mod\'elisation, Universit\'e de Cergy-Pontoise, CNRS, UMR 8089\\
2, Avenue Adolphe Chauvin, 95302 Cergy-Pontoise Cedex, France}




%


\date{\today}

\begin{abstract}
It is experimentally observed  and theoretically proved  that the distance between topmost layers of  a metal surface
has a  contraction. However, well-known potentials such as Lennard-Jones and Morse potentials
 lead to an expansion of the surface inter-layer distance. Such simple potentials therefore cannot be used to study metal surface relaxation.
In this paper, extensive Monte Carlo simulations are used to study the silver (111) surface with both the Gupta potential (GP) and the Embedded Atom Method (EAM) potential.
Our results of the lattice relaxation at the (111) surface of silver show indeed a contraction for both potentials at low temperatures in agreement with experiments and early theories.  However at higher temperatures, the EAM potential yields a surface melting at  $\simeq 700$ K very low with respect to the experimental bulk melting at $\simeq 1235$ K while the GP yields a surface melting at  $\simeq 1000$ K  closer to the bulk one.  In addition, we observe with the EAM potential an anomalous thermal expansion, i. e. the surface contraction becomes a surface dilatation with respect to the bulk, at  $\simeq 900$ K.  The Gupta potential does not show this behavior.We compare our results with different experimental and numerical results.


\begin{description}
\item[PACS numbers:05.10.Ln ; 64.70.dj 	; 05.70.Np ]
\item[Structure]
\end{description}
\end{abstract}

\maketitle


\section{\label{sec:level1}INTRODUCTION}

It is well known that surface atoms of a material have a behavior different from that of bulk atoms, mainly because of the lack of neighbors and the surface geometry.
Different kinds of surface behavior can occur according to the nature of the
material and the surface orientation.
The case of metallic materials has been well studied theoretically \cite{Jona1978,Friedel1976,Gupta1981,Barnett1983,Landman1980,PerdewJ.1982,Lewis1994} and by means of different experimental techniques \cite{Statiris1994,Soares2000,Joost1987}.

The contraction of the lattice spacing is experimentally observed by Medium Energy Ion Backscattering (MEIS) \cite{Statiris1994}, Low-Energy Electron Diffraction (LEED) \cite{Soares2000} or X-ray scattering \cite{Botez2001}.
Other methods such as elastic He scattering and electron energy-loss spectroscopy have also been used. In a theoretical point of view, Gupta has shown analytically that for classical pairwise potentials (Lennard-Jones, Morse, ...), inter-layer distance near the surface have an expansion.
He has also shown that the Tight-Binding-Potential (TBP) and the so-called Gupta potential (GP) lead to a contraction of inter-layer distance at metallic surfaces.

Despite  the important number of experimental and theoretical techniques  used to observe this phenomenon,  there  exists an unsolved question on how the contraction evolves with increasing temperature.
There are two contradictory answers in the literature:   X-ray scattering \cite{Botez2001} and LEED \cite{Soares2000} as well as molecular dynamics (MD) simulations using the Embedded Atom Method (EAM) \cite{Lewis1994,Al-Rawi2000} show
the surface inter-layer distance always smaller than the bulk one at the same temperature, namely surface is contracted, whereas MEIS experiments \cite{Statiris1994} and  {\it ab-initio}
density-functional theory (DFT) calculations \cite{Narasimhan1998} show an anomalous
thermal expansion of the surface at some temperature below the bulk melt.
Facing this long-standing unsolved question, we wanted to carry out  a Monte Carlo (MC) study in an attempt to answer that question.
To our knowledge, there is no MC simulations in literature so far about this subject, although some MC simulations have been used to reproduce experimental patterns
 such as surface blocking pattern during scattering process [see Ref. \cite{Joost1987} for the Pb (110) surface].  Given a tremendous number of numerical studies on surface problems,  it is surprising that no MC simulation has been performed so far to see the variation of the surface  inter-layer distance.

The purpose of this paper is thus to investigate by MC simulation the variation of the lattice spacing between the topmost layers of the (111) silver
surface versus temperature.

In order to simulate such a behavior as accurately as possible, we have  considered  potentials
 which describe as well as possible the material.
The EAM potential is often used in MD simulations and especially for the Ag  (111)  surface \cite{Lewis1994,Al-Rawi2000,Kara1997}. Working with this potential allows us to compare our results with  other numerical studies using the same potential. Furthermore, the EAM potential reproduces accurately the bulk melting temperature of Ag. On the other hand, the Gupta potential describes well surface and cluster
behaviors \cite{Diep1989cluster,Shao2005,Garzo2000,Rogan2008,Chen2011}. The melting temperature of bulk Ag is also well reproduced with this potential.
That was the reason why the two potentials GP and EAM have been used since many years  to simulate silver material and
other metallic crystals.
However, as will be seen below, the two potentials, although yielding the same result for low-temperature surface contraction, give different results at higher temperature concerning the surface contraction and the surface melting.


In Section \ref{sec:level2}  we recall essential properties of the two potentials with their sets of parameters and we briefly describe our algorithm.
 The results obtained by our computation are shown in Section \ref{sec:level3}.
 Concluding remarks are given in Section \ref{sec:level4}.

\section{\label{sec:level2}MODEL AND MONTE CARLO METHOD}

\subsection{Potentials}

\subsubsection{Gupta potential}
In computer simulation of metals, the Gupta  potential is one of the most used semi-empirical potentials. There is a multiple reason for this success but the main ones are the accuracy
of its results for metals and its easy and quick implementation.
The Gupta potential is derived from the Gupta's expression of the cohesive energy of the bulk material and is based on the second-moment approximation of the electron density of states
 in the tight-binding theory. It includes implicitly some many-body interactions.
The expression of the potential is the following:

\begin{equation}{\label{eq1}}
V(r) = E_n\sum_i^N{{\left[A\sum_{{i \ne j}}^n{e^{-p\left(\frac{r_{i,j}}{r_0}-1\right) }} - \sqrt{\sum_{i \ne j}^N{e^{-2q\left(\frac{r_{i,j}}{r_0}-1\right)}}}\right]}}
\end{equation}
where $A$ is a constant given in eV, $r_0$  the equilibrium nearest-neighbor (NN) distance in the bulk metal,
$p$  the repulsive interaction range and $q$ the attractive one. $r_{i,j}\ $
is the distance between the atoms $i$ and $j$ and  $E_n\ $ an energy constant which depends on the size of the system.
In Eq. (\ref{eq1}), the first part gives is a Born-Mayer pairwise repulsion energy term  and the second part is the  many-body attractive contribution.

The parameters used in this work are reported in Table \ref{tab:table1}.

\begin{table}[h]
\caption{\label{tab:table1}%
Gupta parameters for silver}

\begin{ruledtabular}
\begin{tabular}{ll}
\textrm{Parameter}&
\textrm{Value\footnote{Reference \cite{Shao2005}}}\\
\colrule
$A$ \ (eV) & 0.09944\\
$p$  & 10.12\\
$q$  & 3.37\\
$r_0 $(\AA) & 2.892\\
$E_n$ & 2.52\footnote{Fitted with experimental bulk melting temperature in  \cite{BocchAg} } \\
\end{tabular}
\end{ruledtabular}
\end{table}

\subsubsection{EAM potential}
 Several authors \cite{Murray1983,Foiles1986,Zhou2001} have proposed a method based on density-functional ideas called  EAM. We used a version of the EAM potential given in Ref. \cite{Zhou2001}.
The parameters of the EAM potential used in this work for silver are reported in Table \ref{table2}.
In the EAM potential, the total potential energy is given by:

\begin{equation}{\label{eq2}}
E_p = \sum_i\left[F_i\left(\rho_i\right) + \frac{1}{2}\sum_{j \ne i}\phi\left(r_{ij}\right)\right]
\end{equation}
where $\phi_{ij}\ $ represents the pair energy between two atoms $i$ and $j$ at the distance $r_{ij}$. $F_i\ $ is the embedding energy function which represents the energy to embed
an atom into a local site with electron density $\rho_i$. The electron density $\rho_i\ $ has the following expression:

\begin{equation}{\label{eq2}}
\rho_i = \sum_{j\ne i}f\left(r_{ij}\right)
\end{equation}
with
\begin{equation}{\label{eq3}}
f\left(r_{ij}\right) = \frac{f_e\exp\left[-\beta\left(\frac{r_{ij}}{r_e}-1\right)\right]}{1+\left(\frac{r_{ij}}{r_e}-\lambda\right)^{20}}
\end{equation}
where $r_e$, $f_e$, $\beta$ and $\lambda$ are constant parameters which are given in Table \ref{table2}.
The embedding function has the following form:

\begin{equation}
F\left(\rho\right) = \left\{\begin{array}{ll}
\sum_{i=0}^{3}F_{n_i}\left(\frac{\rho}{\rho_n}-1\right)^i &  \rho<0.85\rho_e,\\
\sum_{i=0}^{3}F_{i}\left(\frac{\rho}{\rho_e}-1\right)^i &  0.85\rho_e\le\rho<1.15\rho_e,\\
F_{e}\left[1-\ln\left(\frac{\rho}{\rho_e}\right)^\eta\right]\left(\frac{\rho}{\rho_e}\right)^\eta &  1.15\rho_e \le \rho
                    \end{array}\right\}
\end{equation}
The pair energy expression between atoms $i$ and $j$ is :

\begin{equation}{\label{eq4}}
\phi\left(r_{ij}\right)=\frac{A\exp\left[-\alpha\left(\frac{r_{ij}}{r_e}-1\right)\right]}{1+\left(\frac{r_{ij}}{r_e}-\kappa\right)^{20}}-
\frac{B\exp\left[-\beta\left(\frac{r_{ij}}{r_e}-1\right)\right]}{1+\left(\frac{r_{ij}}{r_e}-\lambda\right)^{20}}
\end{equation}
All constant parameters in the above expressions such as $F_{n_i}$ etc. are listed in Table \ref{table2}.  Note that MD simulations using EAM potential yield the bulk melting temperature at 1170 K for Ag \cite{Foiles1986} which is to be compared to the experimental value of 1235 K \cite{Touloukian1975}.

\begin{table}[h]
\caption{\label{table2}%
EAM parameters for silver}

\begin{ruledtabular}
\begin{tabular}{ll|ll}
\textrm{Parameter}&
\textrm{Value\footnote{Reference \cite{Zhou2001}}}&
\textrm{Parameter}&
\textrm{Value}\\



\colrule\\
$r_e $& 2.891814&$F_{n_1}$ & -0.221025\\
$f_e$ &1.106232 &$F_{n_2}$ & 0.541558\\
$\rho_e $&15.539255 &$F_{n_3}$ & -0.967036\\
$\alpha$ &7.944536 &$F_0$ & -1.75\\
$\beta$ &4.237086 &$F_1$ & 0\\
$A$ &0.266074 &$F_2$ & 0.983967\\
$B$ &0.386272 &$F_3$ & 0.520904\\
$\kappa$ &0.425351 &$\eta$ & 1.149461\\
$\lambda$ &0.850703 &$F_e$ & -1.751274\\
$F_{n_0}$ &-1.729619 \\
%
\end{tabular}
\end{ruledtabular}
\end{table}

\subsection{The algorithm}
The complete description of the algorithm can be found in Ref. \cite{Bocrares}.  For the present surface problem, we have used the the periodic boundary
conditions (PBC) in two directions $x$ and $y$.
Let us briefly recall the main steps of the algorithm.
The step that requires a long CPU time is the construction of  the PBC table which contains neighbors of
each atom of the simulation sample. The reader is referred to Ref. \cite{Bocrares} for details.
Note that to apply correctly the PBC we have to take into account instantaneous fluctuations of the sample
size in each direction at every MC step.

\begin{figure}[!h]
 \includegraphics[width=8.5cm,height=5cm]{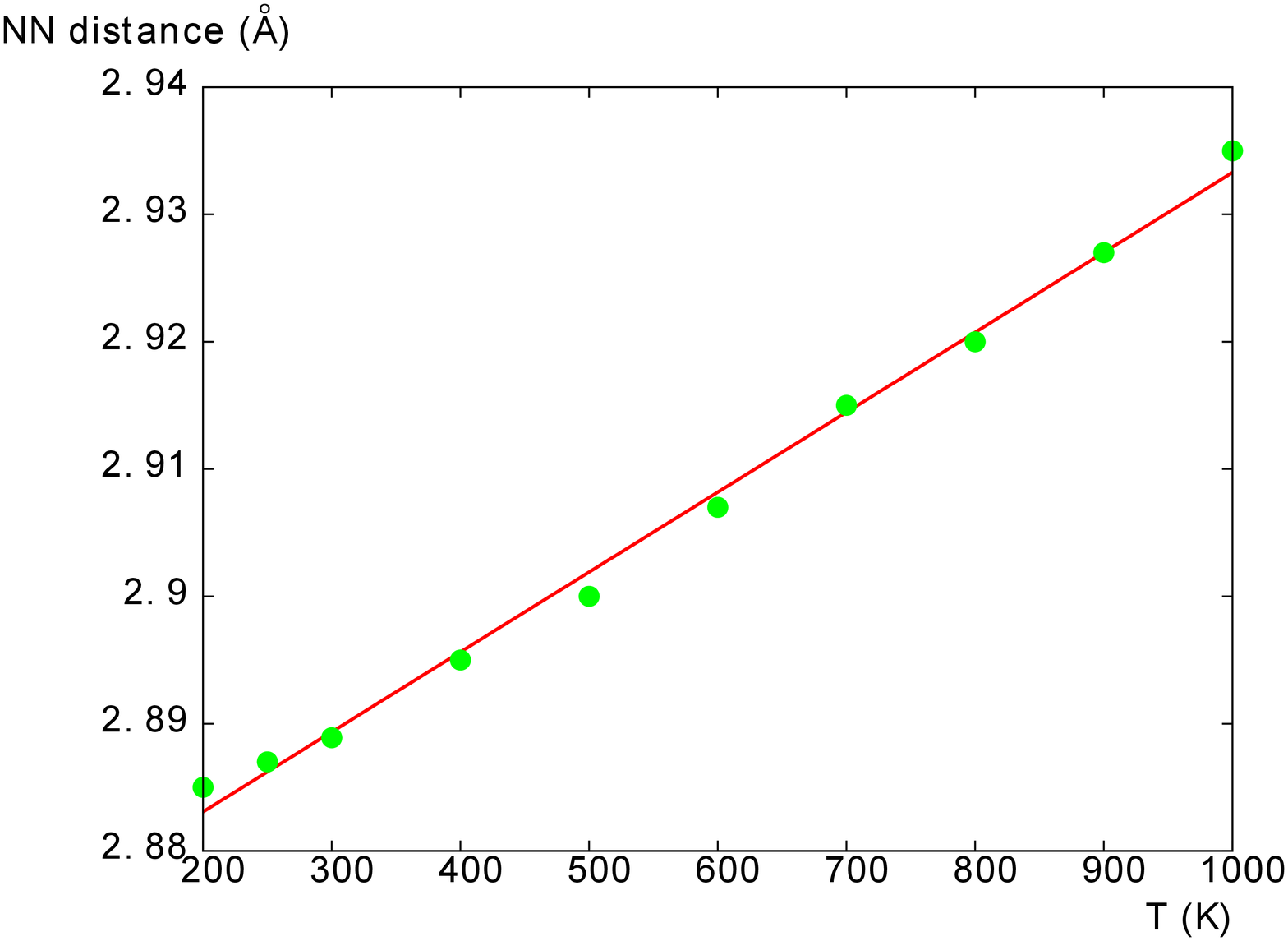}
\caption{\label{avt} (Color online) Experimental data by X-rays diffraction measurements(green circles) from
Ref. \cite{Touloukian1975} and the least square linear fit (red line).}
\end{figure}

\begin{figure}[!h]
 \includegraphics[width=8cm,height=4cm]{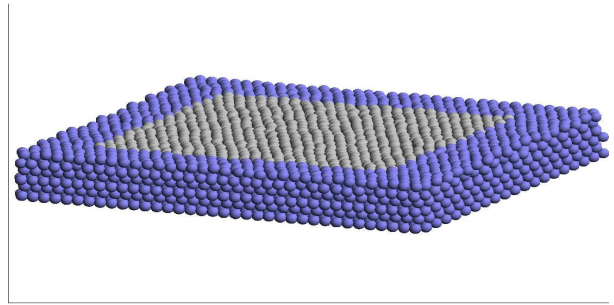}
\caption{\label{2layers}(Color online) Side view of a silver sample, with periodic boundary conditions, after several millions of MC steps at 300 K. Only atoms of  the three topmost layers are allowed to move. See text for
explanation.}
\end{figure}

Using the least square approximation for experimental data of the NN distance
 measured for a bulk sample taken from Ref. \cite{Touloukian1975} we obtain  the linear thermal expansion  $d_{NN}(\textrm{\AA})=\alpha_L T+2.87053$  with the  coefficient
 $\alpha_L=6.27617\times10^{-5}\ $.
In Fig. \ref{avt} we have plotted  experimental data taken from Ref. \cite{Touloukian1975} and the least square approximation used in this paper.
The lattice constant at a given temperature is now an input data. In our calculations, we begin
at $T=200$ K in order to stay in the classical limit. Indeed, as we can see in Ref. \cite{Haug2008} that
at lower $T$, the thermal expansion coefficient  does not have the same linear behavior
below and above 200 K.
At each  temperature, we start the simulation with the lattice constant taken from Fig. \ref{avt}.  In doing so, we favor a fast convergence to equilibrium. This was what we observed in test runs.
Only the surface atoms and an adjustable number of layers beneath the surface are moving. It allows us to choose a number of  moving layers to produce precise results while saving CPU time. Of course, we have to make sure
that final results do not depend on the choice of the number of moving atoms (see an example in Fig. \ref{2layers}).


Note that by using the correct value of the lattice constant at each temperature for interior layers
far from the surface, we take into account some temperature effects but we neglect local fluctuations
in those far layers. The effect of such approximation has been tested: the result confirms that this does not affect
our conclusion.
As seen below, in most simulations which have been made, atoms are allowed to  move in the first three layers.
We have observed that the contraction
of the surface lattice spacing  disappears rapidly at a few layers from the surface. This approximation
is justified by tests  we have made and by experimental observations and theoretical calculations mentioned above.

Note that we use the Metropolis updating criterion. At a given temperature, we equilibrate the system before computing the averages of
 quantities needed for analyzing the system behavior as functions of temperature.

\section{\label{sec:level3}RESULTS AND DISCUSSIONS}

In this section we describe our MC results for the (111) surface of silver.
We have studied a system of 256 atoms per layer in which atoms in the first three layers are
allowed to relax at each MC step.  From the fourth layer inward, the atoms dilate uniformly
with temperature using data of Fig. \ref{avt}.  For comparison, we have also considered a system of 100 atoms per layer with 3 moving layers.  As it turned out, the thermal relaxation of
the first moving layers is not significantly different for these two surface sizes.
Therefore, we show below our results obtained with the surface size of 256 atoms on the top of five beneath layers (6 layers in all) with the PBC in the $x$ and $y$ directions.  Usually, our simulations are carried out over 5 to 6 millions of MC steps per atoms.  We discarded the two first millions during equilibrating time and averaged physical quantities over the following 2-3 millions of MC steps.

\subsection{Surface melting}

The first step is the monitoring of the surface stability versus temperature. Of course,
our calculations have been carried out for temperatures below the bulk melting temperature ($T_m$)
 because we use the temperature-dependent bulk lattice spacing below $T_m$ shown in Fig. \ref{avt}
 for bulk atoms.
 Note however that very close to $T_m$, the thermal coefficient does not follow the linear behavior.
We have therefore heated the system from the ground state to temperatures
between 200 K and 1200 K, just below $T_m$,  with a step of 50 K.

 \begin{figure}[!h]
  \includegraphics[width=8.5cm,height=5.2cm]{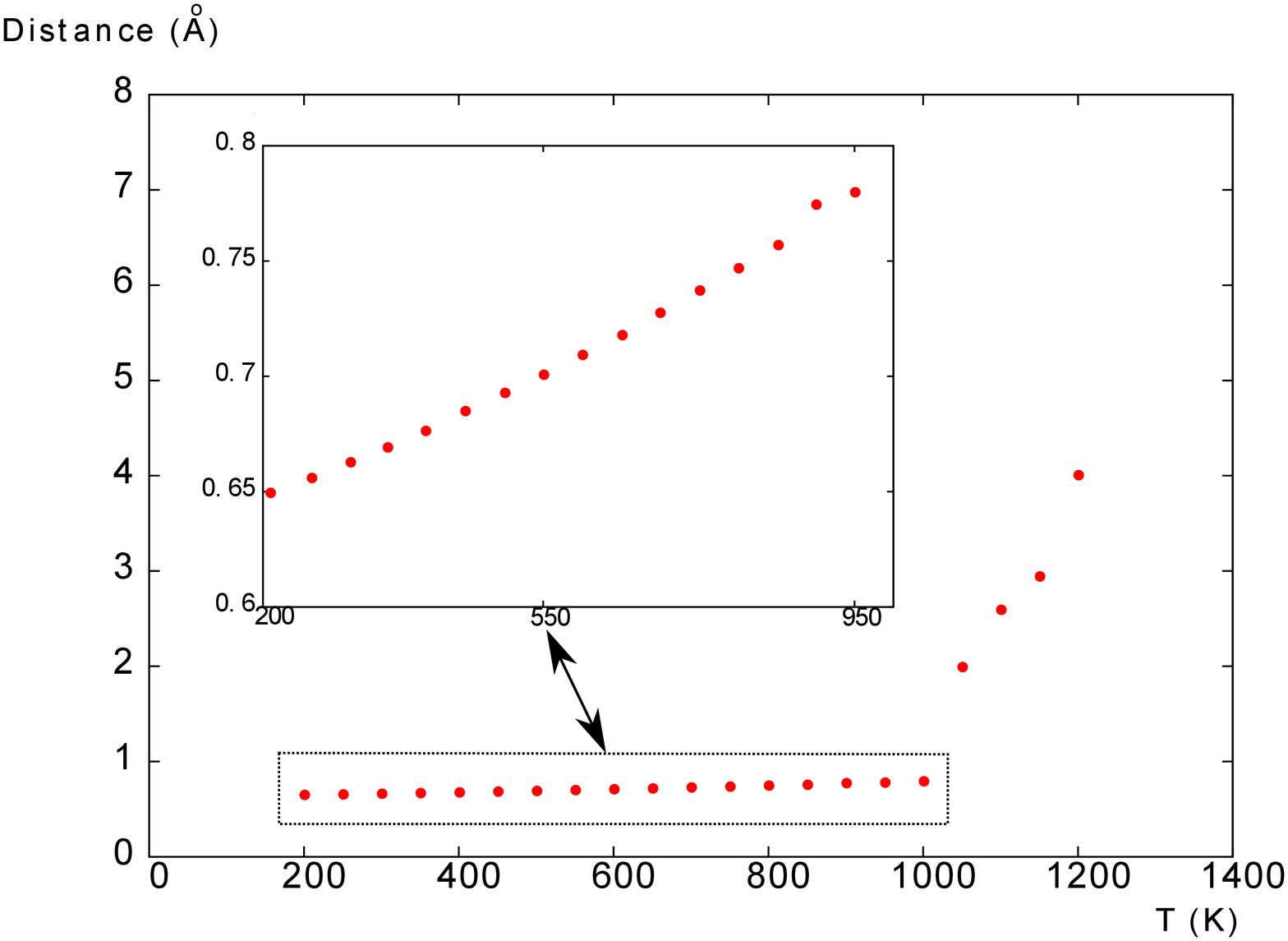}
 \caption{\label{evt_G}(Color online) Mean displacement amplitude (in \AA) between atoms
 and their perfect ground-state lattice positions computed with Gupta potential.}
 \end{figure}

The mean displacement amplitude allows us to establish the validity domain of temperatures. This quantity is
computed as follows : at each MC step, we compute a spatial average of all
separation distances between atoms with their
ground-state node positions and then we compute its thermal average through MC steps.
 As we can see in Fig. \ref{evt_G}, with the Gupta potential,  after 1000 K the atoms undergo a sudden
 jump indicating a phase change: the surface becomes disordered (melted) between 1000 K and 1050 K. This value from our MC simulation using Gupta potential is very close to the value 1100 K obtained from MD simulation using the EAM potential \cite{Al-Rawi2000}.
 A study is needed to determine the precise surface melting temperature but it is not the aim to search
  for such a precision of this paper.

\begin{figure}[!h]
\includegraphics[width=8.5cm,height=5.2cm]{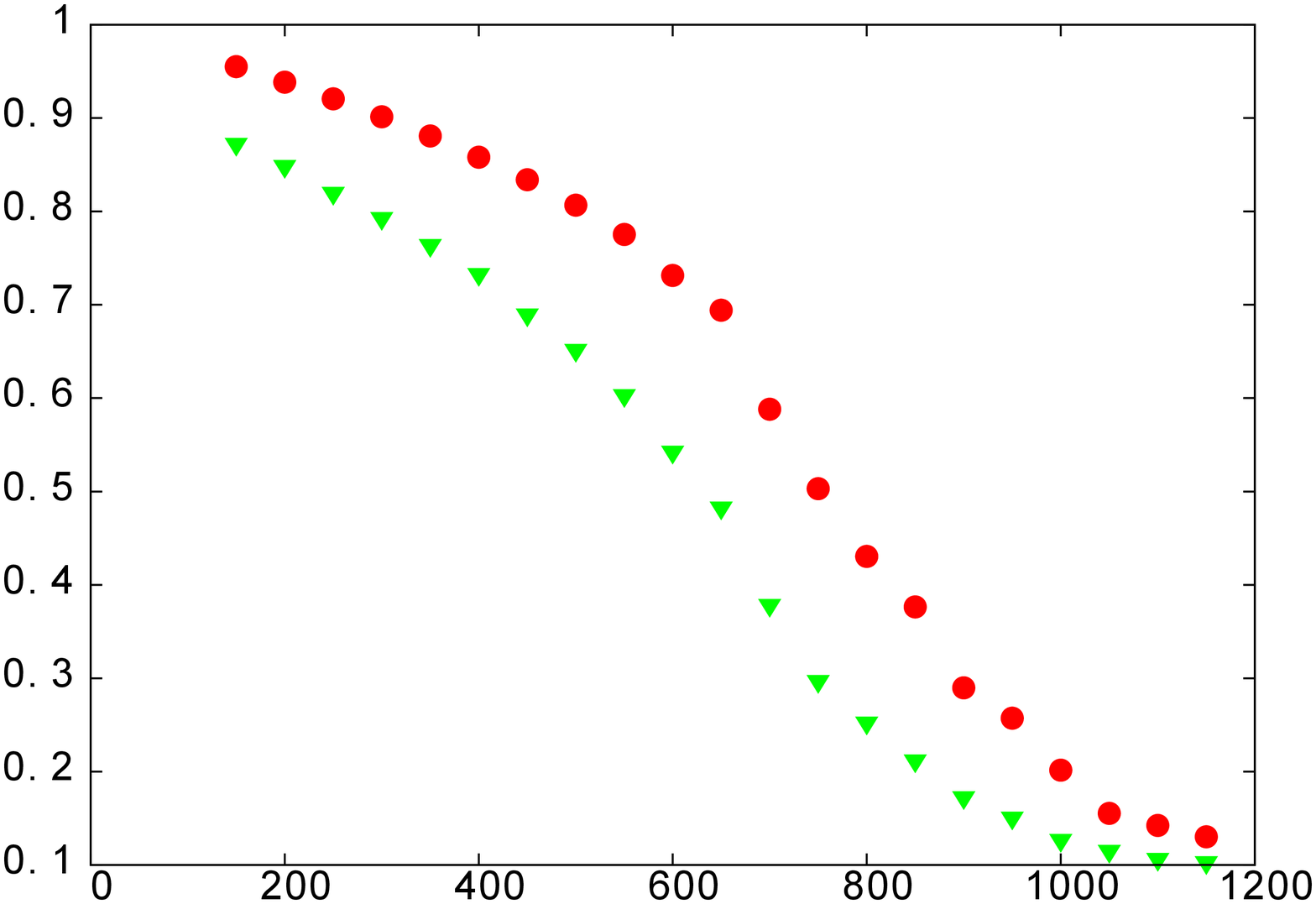}
\caption{\label{structure_factor}(Color online) Structure factor $S_{\vec K}$ (green triangles) and  $O_6$ order parameter (red circles) of the first layer versus temperature for the EAM potential.}
\end{figure}

For the EAM potential, we have obtained  a result different from that of MD simulation: the surface melting temperature takes place at about 700 K, while MD simulation fonds it at 1100 K as we  mentioned above.
In order to see clearly this surface melting, we have computed with
the EAM potential the structure factor for the topmost layers.
The structure factor $S_{\vec K}$ is computed as follow:

\begin{equation}
S_{\vec K} =\frac{1}{N_l}\left < \left |\sum_{j=1}^{N_l}e^{i\overrightarrow{K}.\overrightarrow{d_j}}\right |\right >
\end{equation}
where $\overrightarrow{d_j}$ is the position vector of an atom in the layer, $N_l$  the number of atoms in a layer and $\overrightarrow{K}$  the reciprocal lattice vector which has the following coordinate (in reduced units):
$2\pi\left(1;-\sqrt{3};0\right)$.  The angular brackets $<...>$ indicate thermal average taken over MC run time.
The above  "order parameter", which allows us to monitor the long-range surface order,
is plotted for the surface layer in Fig. \ref{structure_factor}.
 As we can see, the long-range order is lost  at $\simeq 700$ K.
 In order to investigate in more details the surface melting, we have also computed the $O_6$ order parameter introduced in Ref. \cite{Carnevali1987}. This order parameter describes the short-range hexatic orientational
order of the surface which is defined as follows:

\begin{equation}
O_6 =\frac{\left|\sum_{jk}W_{jk}e^{i6\Theta_{jk}}\right|}{\sum_{jk}W_{jk}}
\end{equation}
with
\begin{equation}
W_{jk} = e^{- \frac{\left(z_j-z_k\right)^2}{2\delta^2}}
\end{equation}
where the sum runs over the NN pairs and $\Theta_{jk}$ is the angle which the $j-k$ bond, when projected on the $xy$ plane, forms with the $x$ axis. The $\delta$ parameter is taken as one-half the average inter-layer spacing.
The weighting function, $W_{jk}$, allows us to differentiate the "non coplanar" and the "coplanar" neighbors. With a coplanar neighbor, the weighting function takes a maximum value.
We have to calculate the spatial average of $O_6$ taken over all atoms of the surface layer and then calculate its thermal average over MC run time. We plot the so-averaged $O_6$ parameter  and the $S_{\vec K}$ structure factor versus temperature in Fig. \ref{structure_factor}.

We observe that the long-range and short-range orders are lost at  the same temperature: the melting of the surface occurs at around 700 K where both order parameter change their curvature.

\begin{figure}[!h]
 \includegraphics[width=4.2cm,height=2.6cm]{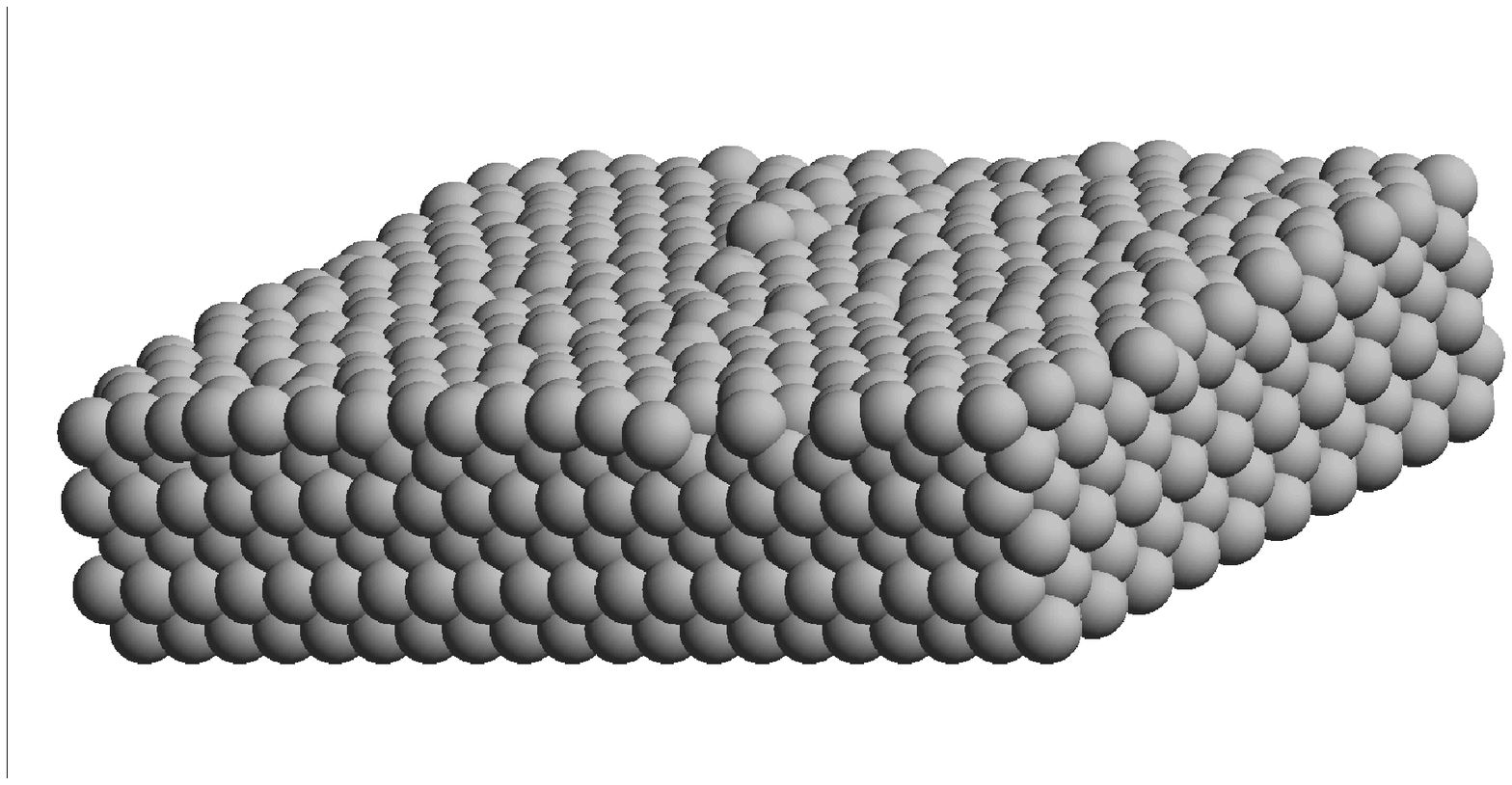}\hfill
 \includegraphics[width=4.2cm,height=2.6cm]{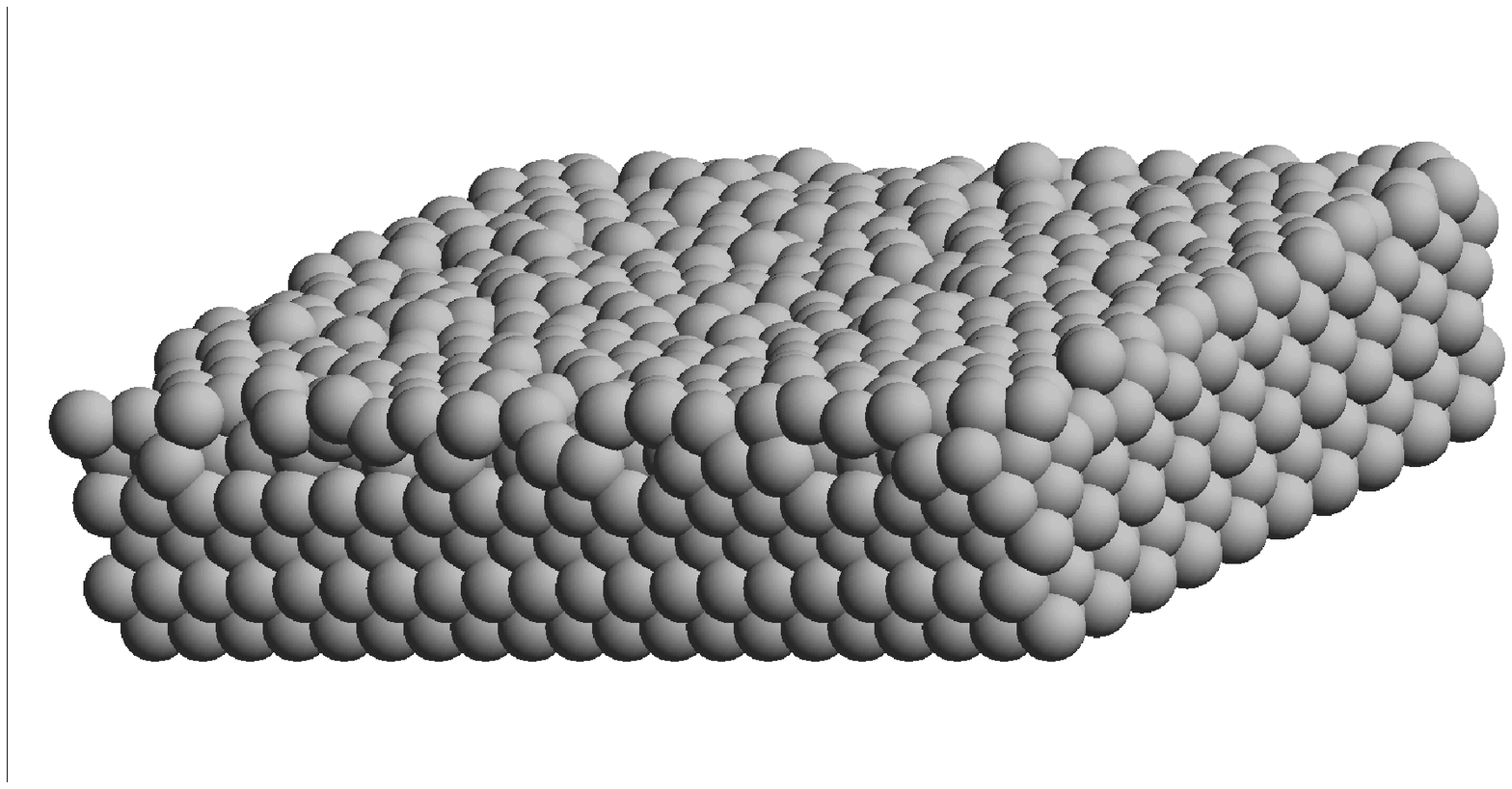}
\caption{\label{pictureseam}Pictures of the system for two different temperatures computed with the EAM potential: 550 K (left) and 650 K (right). Only three layers are allowed to relax.}
\end{figure}

\begin{figure}[!h]
 \includegraphics[width=4.2cm,height=2cm]{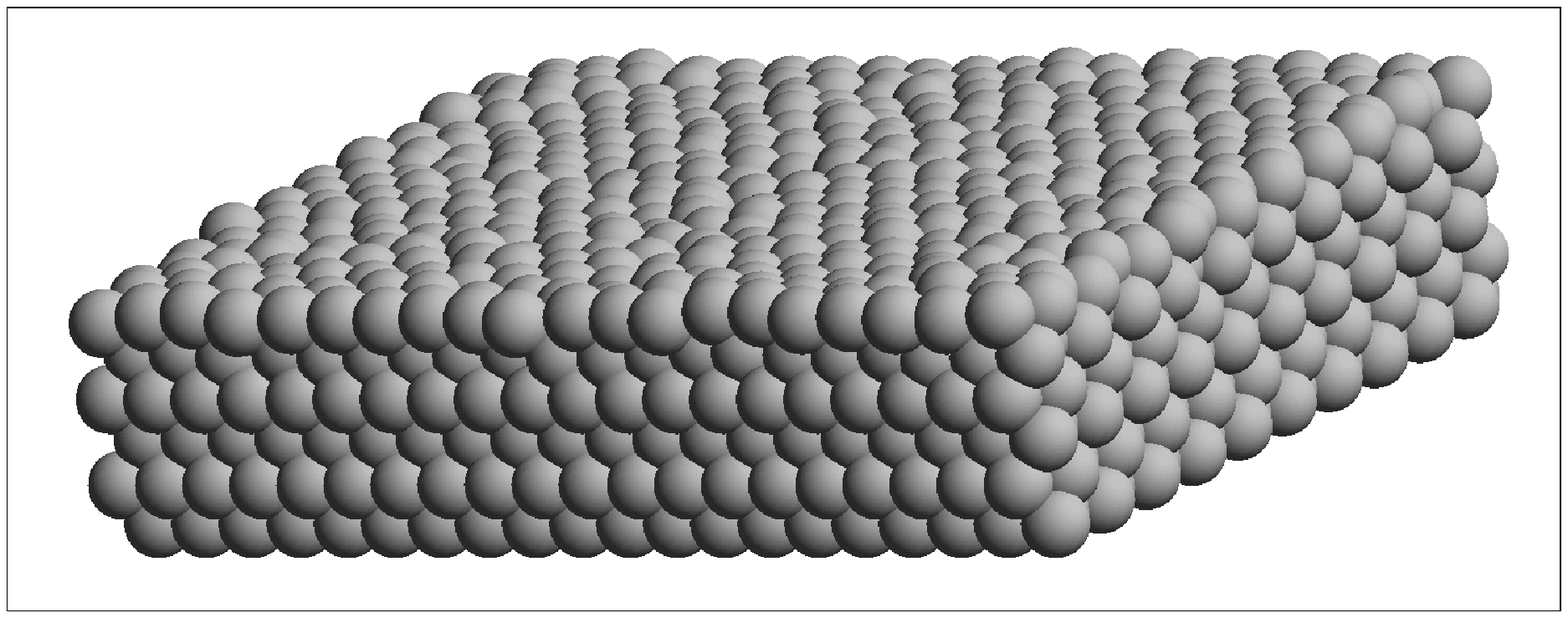}\hfill
 \includegraphics[width=4.2cm,height=2cm]{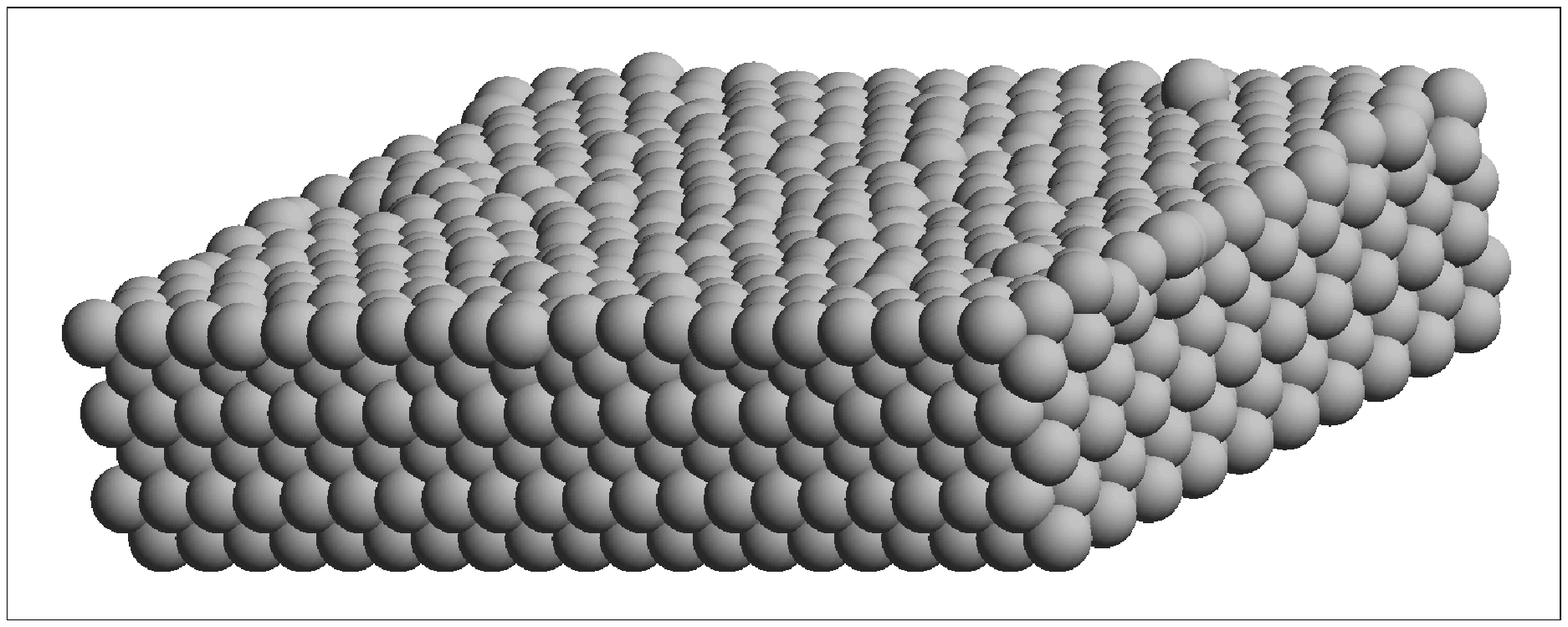}
\caption{\label{picturesgupta}Pictures of the system for two different temperatures computed with the Gupta potential : 950 K (left) and 1050 K (right).Only three layers are allowed to relax.}
\end{figure}

Snapshots of the system
at two temperatures close to the surface melting are shown in Figs. \ref{pictureseam}
and \ref{picturesgupta}, for the two potentials.

The difference in the surface melting temperature  for the two potentials is surprising since both give   the same bulk melting temperature.  One possible explanation comes from the construction of their set of parameters:  the Gupta potential was fitted in Ref. \cite{BocchAg} with the bulk melting temperature, while the EAM potential is fitted to reproduce physical quantities such as the lattice constant, the elastic constants $C_{11},C_{12}\ $and $C_{44}\ $ and
sublimation energy, etc. but not with the bulk melting temperature. A consequence is that surface atoms in the Gupta case are more strongly attached to the interior part, so that we need a much higher temperature to make the surface melt. We believe that the high "stability" of surface atoms is the reason why the Gupta potential was very  suitable for the study of clusters of very few number of atoms \cite{Diep1989cluster}.

For  both potentials, a contraction of
the lattice spacing between two topmost layers occurs but its temperature dependence is different in the two cases, as will be shown below.

\subsection{Surface contraction}

\begin{figure}[!h]
 \includegraphics[width=8cm,height=5.2cm]{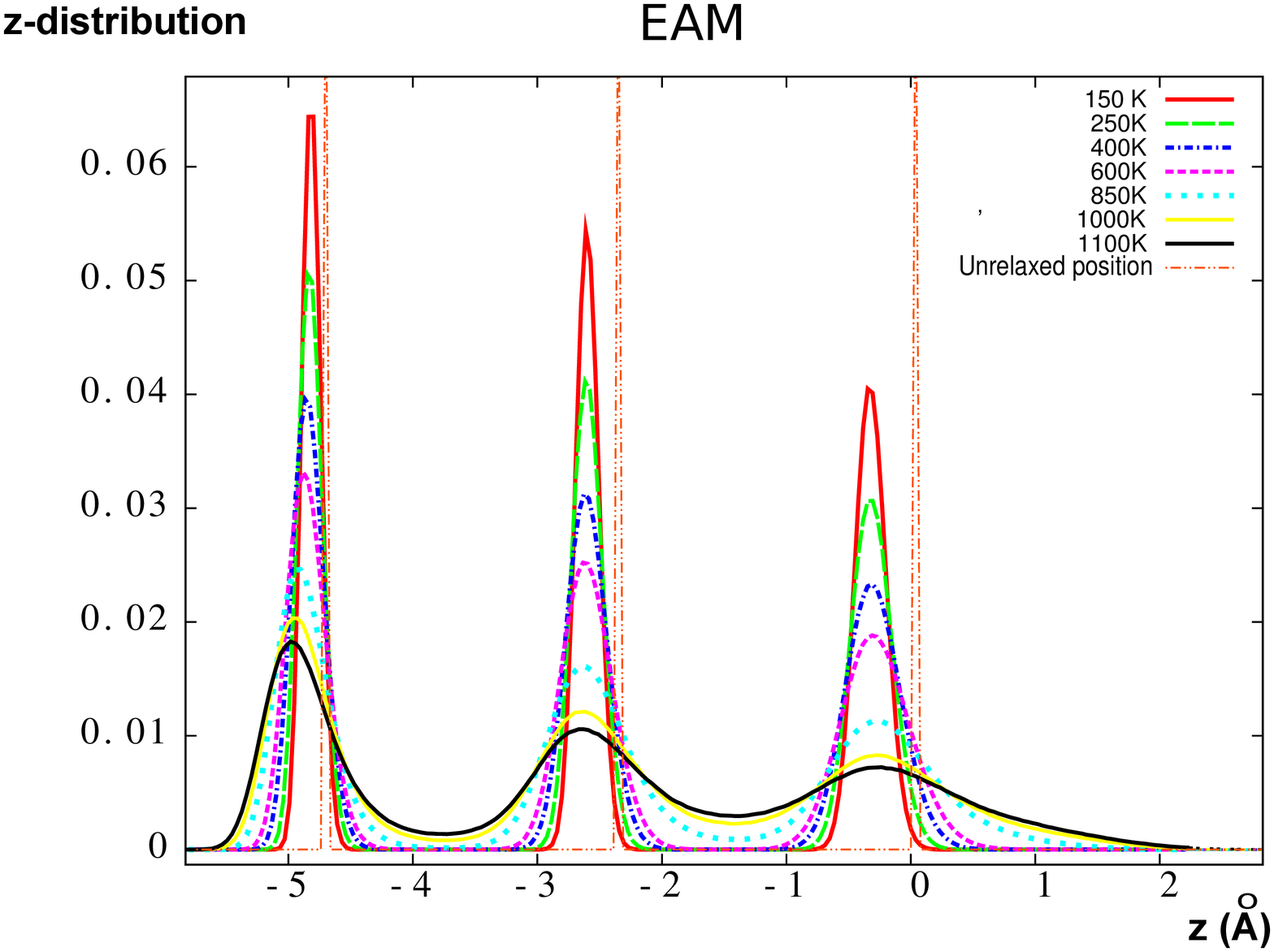}\vfill
 \includegraphics[width=8cm,height=5.2cm]{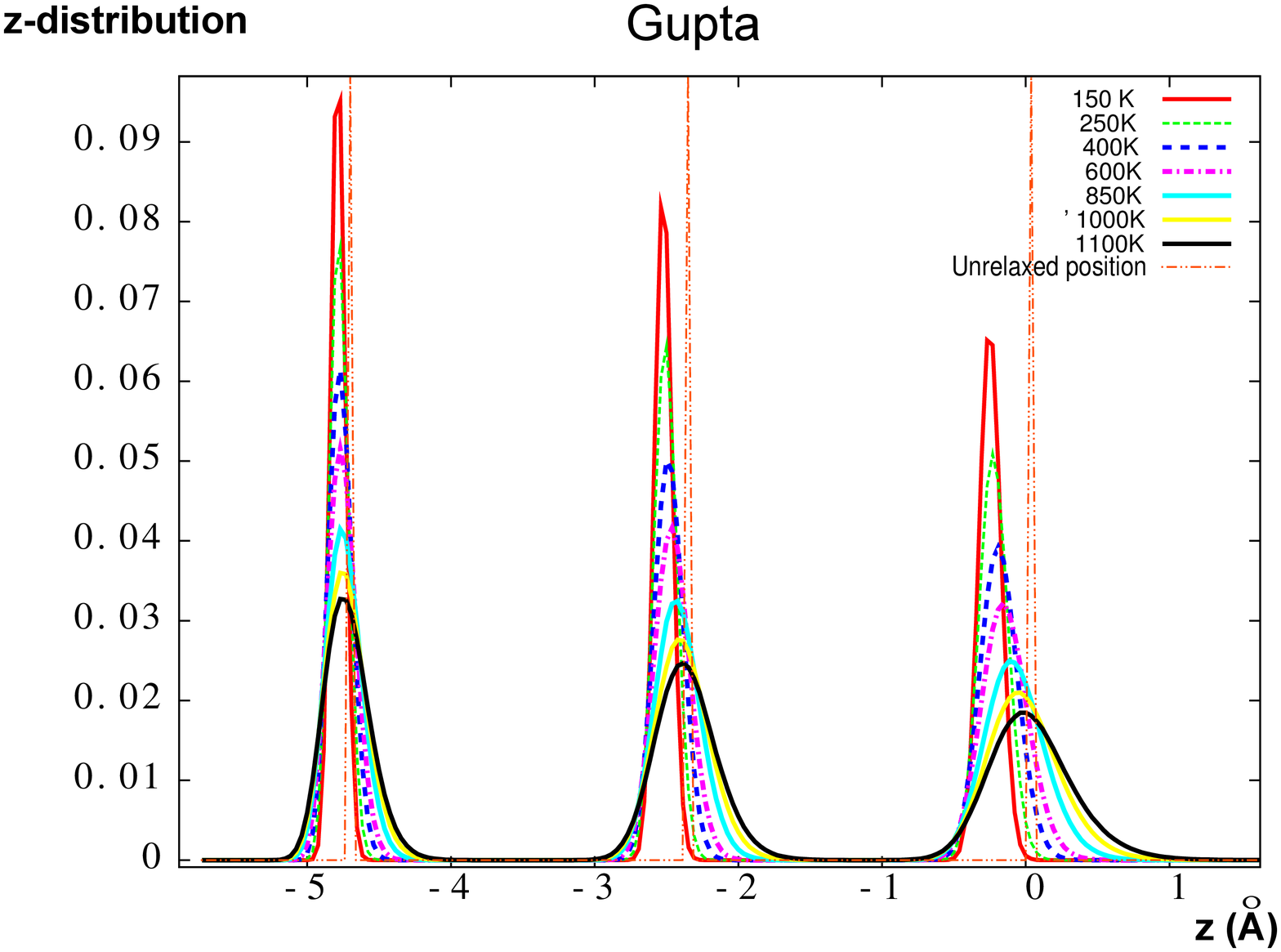}
\caption{\label{distri_z_gupta}(Color online) $z$-position distribution at different temperatures calculated with EAM and Gupta potentials. The surface at its non relaxed position corresponds to $z$=0.}
\end{figure}

In order to see the variation of the inter-layer distance at the surface, we have plotted in Fig.  \ref{distri_z_gupta} the $z$-position distribution obtained  for the
two potentials EAM and GP. As we can see, when the temperature increases,
the three peaks which represent the $z$ coordinates of the three moving layers, are shifted to
the left, indicating a  contraction toward the bulk. Note that the two potentials lead to the same behavior.

In these two $z$-axis distribution functions, we can see that at a given $T$, the deviation of the topmost layer from its perfect position is largest. This deviation becomes smaller for the next two layers and it should disappear at a few layers from the surface. This observation justifies our approximation to let only  the three topmost layers to relax.
Previous theoretical results and experimental data also give support to our hypothesis.

\begin{figure}[!h]
 \includegraphics[width=8.5cm,height=5.2cm]{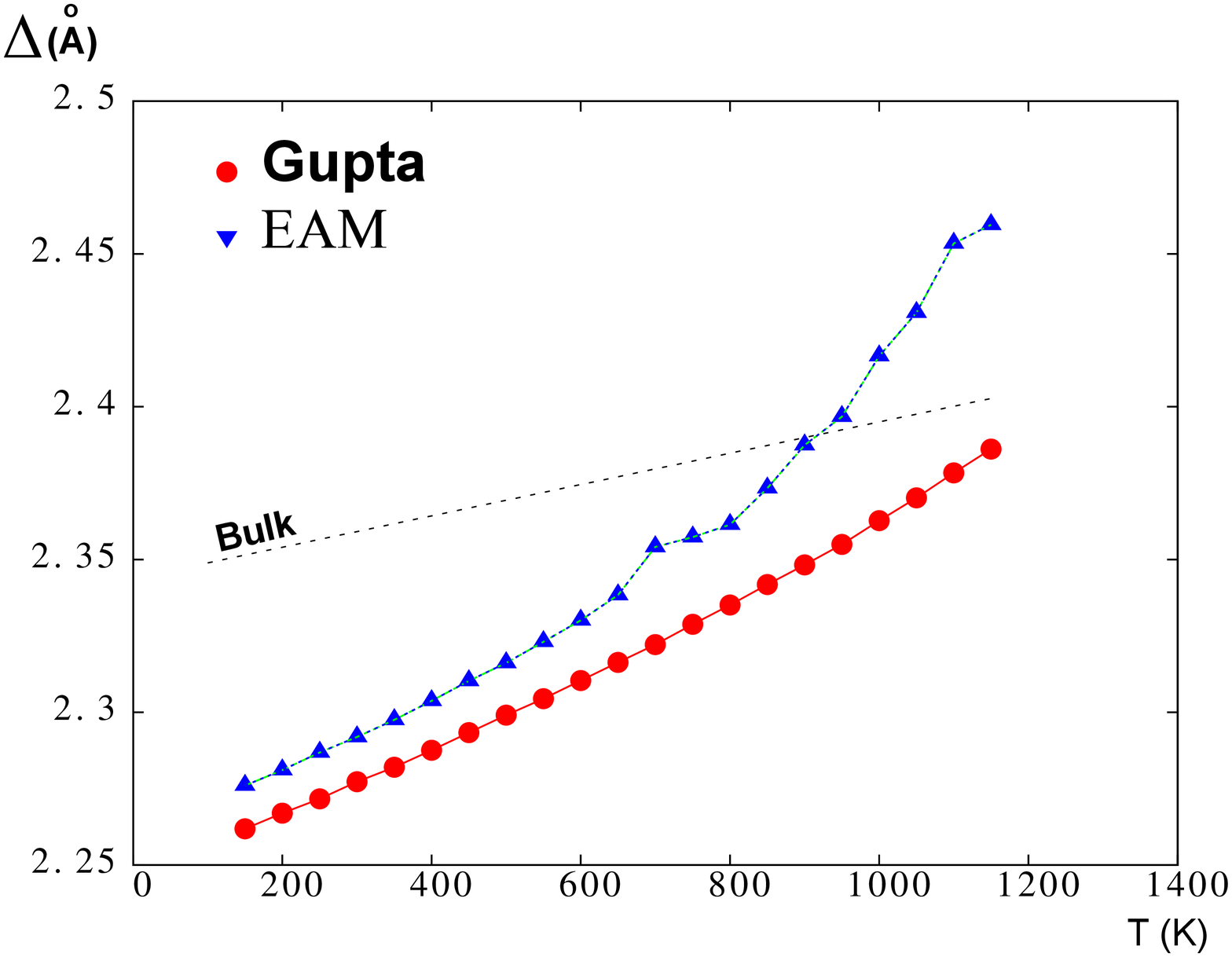}
\caption{\label{12interlayer}(Color online) Inter-layer $\Delta_{12}\ $ computed with Gupta  and EAM potentials. Dashed line represents the bulk inter-layer distance. See text for comments.}
\end{figure}

In order to examine closely  the inter-layer contraction, we have plotted in Fig. \ref{12interlayer} the inter-layer distance $\Delta_{12}\ $ between layer 1 and 2, at different temperatures.
This quantity is determined by averaging over the corresponding peaks during MC simulation time.
We only focus our attention on $\Delta_{12}$ since for this quantity there is a general agreement between experiments and between experiments and theories. For inner  layers, there is no such agreement between experiments and theories, and even between experiments, depending on the technique used: LEED \cite{Soares2000}, MEIS \cite{Statiris1994}, ...

Let us comment the results shown in Fig.  \ref{12interlayer}. Several remarks are in order:

(i) At low temperatures, for both potentials, $\Delta_{12}\ $ has a contraction of about  $2.5\%\ $ at 300 K, compared to a contraction of $0.5\%$ with MD simulations using EAM potential \cite{Lewis1994,Al-Rawi2000}. Experimental contraction of $2.5\%$ was observed in Ref. \cite{Statiris1994}.

(ii) The distance $\Delta_{12}\ $ increases  with increasing temperature: in the case of EAM potential, $\Delta_{12}\ $ crosses the bulk limit at $\simeq 900$ K, indicating a surface expansion, in agreement with experimental results in Ref. \cite{Statiris1994}.  This "expansion" was not found by MD simulations with EAM potential (see Refs. \cite{Lewis1994,Al-Rawi2000}) up to 1100 K and in LEED \cite{Soares2000}. A surface expansion was observed  in an {\it ab-initio} DFT calculation \cite{Narasimhan1998}.

(iii) As we can see in Fig. \ref{12interlayer}, MC results with GP does not show the above-mentioned anomalous expansion up to 1150 K slightly above the surface melting temperature observed above with this potential.
We will give below an explanation about this point at least from our MC results.


 As said above, the contraction of the first inter-layer distance $\Delta_{12}\ $ has been observed by both MEIS and LEED {\it at low temperatures}.  For $\Delta_{23}\ $, the distance between the second and third  layers,  these two methods are not in agreement: at low $T$, MEIS finds a little expansion of $\Delta_{23}\ $ while LEED finds a contraction. Here, our computations are in  agreement with LEED results.

Let us discuss about the controversial point on the anomalous expansion of $\Delta_{12}$, namely the transition from contraction to expansion when  $\Delta_{12}$ crosses the bulk line (see Fig. \ref{12interlayer}). As said earlier, this has been experimentally observed by MEIS at about 750 K \cite{Statiris1994}. We have found this with our simulations using EAM potential at $\simeq$ 900 K.  Our expansion is about 5\% at 1150 K, with respect to the bulk value.
The disagreement between our MC result and MD result using the same potential resides not only in the difference of surface melting temperatures (ours is 700 K, the MD one is 1100 K) but also in the existence of the anomalous expansion.

For GP, we do not observe the anomalous expansion up to 11500 K as shown in Fig. \ref{12interlayer}.  So the existence or not of an anomalous expansion depends on the potential at least with our MC results. Let us explain this as follows:

(i) With the EAM potential:  The surface melts at a temperature much lower than that of the anomalous expansion.
As we can see in Fig. \ref{12interlayer}, at around 700 K, the melting of the first layer causes the apparition of a little peak but $\Delta_{12}$ is still smaller than the bulk inter-layer distance.
The fact that the anomalous expansion does not occur at the melting temperature of the first layer
means that the melted surface is in a "two-dimensional" liquid state in the temperature region between 700 K and 900 K. This liquid surface layer is "detached" from the remaining crystal only at temperatures higher than 900 K.
More accurate experiments of surface melting are desirable to allow a better understanding of this point.

(ii) With the GP: Since the surface melts at $\simeq 1000-1050$ K (see Fig. \ref{evt_G}), if there is an anomalous expansion, this should happen at a higher temperature. Given the fact that with EAM, our MC result indicates the anomalous expansion occurring 200 K above the surface melting temperature, we can imagine the same scenario with the GP: an expansion may be possible at $\simeq 1200-1235$ K, namely at the bulk melting.

To close this section, let us summarize here that (i) the surface melting occurs below the bulk one, but the distance to the bulk melting temperature depends on the potential: the EAM and Gupta potentials give the same bulk melting temperature but different surface melting temperatures (ii) an anomalous dilatation of surface is possible only at a temperature much higher than the surface melting temperature.

\vspace{0.5cm}

\section{\label{sec:level4}CONCLUDING REMARKS}
Monte Carlo simulations with both EAM and Gupta potentials show two different surface melting temperatures for the (111) surface of a silver sample: while both EAM and Gupta potentials give the same bulk melting temperature ($\simeq 1200$ K),  the EAM potential yields a surface melting at $\simeq 700$ K and the Gupta potential shows a surface melting at $\simeq 1000-1050$ K close to the bulk melting.

However, both potentials show a contraction of the topmost inter-layer distance of that surface at low temperatures, in good agreement with experiments and theories.
Our results of the temperature-dependence of the inter-layer contraction indicate a strong potential-dependence: the variation of the contraction with increasing
temperature shows a difference of the two potentials.  Using the EAM potential, our MC simulations show an anomalous thermal expansion, namely the distance between the topmost layers is larger than that between two adjacent  bulk layers.  This can be explained by the fact that the anomalous thermal expansion occurs only at a temperature much higher than the surface melting temperature: the already-disordered surface layer is loosely attached to the crystal. This surface "dilatation" is in agreement with MEIS experiments and {\it ab-initio} DFT calculations.    On the other hand, using the Gupta potential, our results show that the surface inter-layer distance varies with the temperature but it is always smaller than the bulk distance at least up to  temperatures close to the bulk melting.  We attribute the non observation of a  surface dilatation in this case  by the fact that surface melting occurs too close to the bulk melting.  The high surface melting temperature has been found in MD simulations with EAM potential \cite{Al-Rawi2000}.  The existence of surface dilatation has been observed  in  MEIS experiments \cite{Statiris1994} but not in LEED experiments \cite{Soares2000} and in
X-ray scattering \cite{Botez2001}. We think that this difference comes in part from  the difference of surface flatness, surface cleanness (contaminated or not) and experimental conditions. We hope that experimentalists could resolve this important question on anomalous high-temperature behavior of Ag(111).   Experiments to determine the melting temperature of the Ag (111) surface is also needed to choose a potential appropriate for Ag crystal.

%
%


{}

\end{document}